\documentstyle[amssymb,amscd,12pt]{amsart}

\newtheorem{cor}{Corollary}
\newtheorem{lem}{Lemma}
\newtheorem{prop}{Proposition}

\theoremstyle{definition}

\theoremstyle{definition}
\newtheorem{thm}{Theorem}

\theoremstyle{remark}
\newtheorem{rem}{Remark}

\begin{document}

\newcommand{\thmref}[1]{Theorem~\ref{#1}}
\newcommand{\secref}[1]{Sect.~\ref{#1}}
\newcommand{\lemref}[1]{Lemma~\ref{#1}}
\newcommand{\propref}[1]{Proposition~\ref{#1}}
\newcommand{\corref}[1]{Corollary~\ref{#1}}
\newcommand{\remref}[1]{Remark~\ref{#1}}
\newcommand{\nc}{\newcommand}
\nc{\on}{\operatorname}
\nc{\ch}{\mbox{ch}}
\nc{\Z}{{\Bbb Z}}
\nc{\C}{{\Bbb C}}
\nc{\cond}{|\,}
\nc{\bib}{\bibitem}
\nc{\pone}{\Pro^1}
\nc{\pa}{\partial}
\nc{\F}{{\cal F}}
\nc{\arr}{\rightarrow}
\nc{\larr}{\longrightarrow}
\nc{\al}{\alpha}
\nc{\ri}{\rangle}
\nc{\lef}{\langle}
\nc{\W}{{\cal W}}
\nc{\gam}{\bar{\gamma}}
\nc{\Q}{\bar{Q}}
\nc{\q}{\widetilde{Q}}
\nc{\la}{\lambda}
\nc{\ep}{\epsilon}
\nc{\su}{\widehat{\goth{sl}}_2}
\nc{\gb}{\bar{\goth{g}}}
\nc{\g}{\goth{g}}
\nc{\hh}{\bar{\goth{h}}}
\nc{\h}{\goth{h}}
\nc{\n}{\goth{n}}
\nc{\ab}{\goth{a}}
\nc{\f}{\widehat{{\cal F}}}
\nc{\is}{{\bold i}}
\nc{\V}{{\cal V}}
\nc{\M}{\widetilde{M}}
\nc{\js}{{\bold j}}
\nc{\bi}{\bibitem}
\nc{\laa}{\bar{\lambda}}
\nc{\fl}{B_-\backslash G}
\nc{\De}{\Delta}
\nc{\G}{\widetilde{\goth{g}}}
\nc{\Li}{{\cal L}}
\nc{\fp}[2]{\frac{\pa}{\pa u_{#1}^{(#2)}}}
\nc{\Ve}{{\cal Vect}}
\nc{\sw}{\goth{sl}}
\nc{\La}{\Lambda}

\title{Kac-Moody Groups and Integrability of Soliton Equations}
\author{Boris Feigin}
\address{Landau Institute for Theoretical
Physics, Moscow 117334, Russia and R.I.M.S., Kyoto University, Kyoto 606,
Japan}

\author{Edward Frenkel}
\address{Department of Mathematics, Harvard University,
Cambridge, MA 02138, USA}

\date{November 1993\\
      Revised: March 1994, November 1994}

\maketitle
\vspace{10mm}

\noindent{\bf Summary} A new approach to integrability of affine Toda field
theories and closely related to them KdV hierarchies is proposed. The flows
of a hierarchy are explicitly identified with infinitesimal action of the
principal abelian subalgebra of the nilpotent part of the corresponding
affine algebra on a homogeneous space.

\section{Introduction}

Soliton equations describe infinite-dimensional hamiltonian systems. They
are closely related to infinite-dimensional Lie groups and algebraic
curves. These relations account for complete integrability of soliton
equations, i.e. the existence of infinitely many integrals of motion in
involution.

Recently a new insight has been brought into the theory by the observation
that these integrals of motion can be viewed as classical limits of quantum
integrals of motion of certain deformations of conformal field theories
\cite{zam,ey,hollm}.

In our previous works \cite{ff:toda,monte}, using the technique of free
field realization of conformal field theories (cf. \cite{icm} for a
review), we gave a homological construction of quantum integrals of motion,
corresponding to particular deformations. This enabled us to prove the
existence of quantum integrals of motion for those deformations. Tracing
our construction back to the classical limit, we realized that it provides
a new approach to integrability of classical soliton equations, namely,
affine Toda equations. Here we will present this approach. We hope that it
can be applied to other soliton equations as well.

\subsection{} Recall that a Toda field theory can be associated to an
arbitrary affine algebra $\g$ \cite{mop}. Let $a_i, i=0,\ldots,l$, be the
labels of the Dynkin diagram of $\g$, and $(\al_i,\al_j)$ be the scalar
product of the $i$th and $j$th simple roots of $\g$ \cite{kac}. The system
of Toda equations corresponding to $\g$ can be written in the form
\begin{equation}   \label{toda}
\pa_\tau \pa_t \phi_i(t,\tau) = \sum_{j=0}^l (\al_i,\al_j)
e^{-\phi_j(t,\tau)}, \quad i = 1,\ldots,l,
\end{equation}
where each $\phi_i(t,\tau), i=1,\ldots,l$, is a family of functions in $t$,
depending on the time variable $\tau$, and $$\phi_0(t,\tau) = -
\frac{1}{a_0} \sum_{i=1}^l a_i \phi_i(t,\tau).$$

We call local functionals in $u_i(t) = \pa_t \phi_i(t), i=1,\ldots,l$,
which are preserved under the time evolution of the $\phi_i(t)$'s, the
local integrals of motion of the system \eqref{toda}. Recall that a local
functional is a functional of the form
\begin{equation}    \label{lf}
F[{\bold u}(t)] = \int P({\bold u}(t),\pa_t {\bold u}(t),\ldots) dt,
\end{equation}
where $P$ is a differential polynomial in ${\bold u}(t) =
(u_1(t),\ldots,u_l(t))$.

The equation \eqref{toda} can be written in the hamiltonian form:
$$\pa_\tau {\bold u}(t,\tau) = \{H,{\bold u}(t,\tau)\},$$ where $\{\cdot
,\cdot\}$ is a certain Poisson bracket on the space of functionals in
${\bold u}$, and $H$ is the hamiltonian:
\begin{equation}    \label{H}
H = \sum_{i=0}^l \int e^{-\phi_i(t)} dt.
\end{equation}
The space of local integrals of motion is by definition the kernel of the
linear operator $\{ H,\cdot \}$ on the space of local functionals. For
brevity we will call these integrals of motion {\em Toda integrals}.

\subsection{} Instead of working with the space of local functionals in
${\bold u}$, we will work with the algebra of differential polynomials in
${\bold u}$, i.e. the polynomial algebra in variables $u_i^{(n)} = \pa^n
u_i, i=1,\ldots,l; n\geq 0$. We denote this algebra by $\pi_0$. The action
of $\pa$ on $u_i^{(n)}$ can be extended to a derivation of $\pi_0$ by the
Leibnitz rule. The space of local functionals in ${\bold u}$ is then the
quotient of $\pi_0$ by the image of $\pa$ (the subspace of total
derivatives) and constants. We make $\pi_0$ into a $\Z$--graded algebra by
putting $\deg u_i^{(n)} = -n-1$. This induces a $\Z$--grading on the space
of local functionals.

The hamiltonian $H$ of the Toda equation is a sum of $l+1$ terms $\int
e^{-\phi_i(t)} dt$. It turns out that the operator of Poisson bracket with
each of these terms gives rise to a certain derivation $Q_i$ of $\pi_0$.

The crucial observation, which enabled us to describe Toda integrals in
\cite{ff:toda,monte} was that the operators $Q_i$ satisfy the {\em Serre
relations} of the Lie algebra $\g$. In other words, these operators
generate an action of the nilpotent Lie subalgebra $\n_+$ of $\g$ on
$\pi_0$. This allowed us to interpret the space of Toda integrals as the
first cohomology of $\n_+$ with coefficients in the module $\pi_0$,
$H^1(\n_+,\pi_0)$, cf. \cite{ff:toda,monte}. We then computed this
cohomology and found that it is spanned by elements of degrees $-m \in -I$,
where $I$ is the set of exponents of $\g$ modulo the Coxeter number. This
agrees with the results previously established by other methods
\cite{mop,DS,ds,kw,w}.

In the present work we will simplify and investigate further this
construction.

\subsection{} Let $G$ be the Lie group corresponding to $\g$. The upper
nilpotent subgroup $N_+$ of $G$ can be considered as a big cell on the flag
manifold $B_-\backslash G$ of $\g$, where $B_-$ is the lower Borel subgroup
of $G$ (cf. \secref{geocon} for precise definitions). On this big cell,
$\g$ infinitesimally acts from the right by vector fields.

The Lie algebra $\g$ contains a principal Heisenberg Lie subalgebra
$\widehat{\ab} = \ab_+ \oplus \ab_- \oplus \C K$, where $\ab_+$ is the
principal abelian subalgebra of $\n_+$, and $\ab_-$ is its opposite with
respect to a Cartan involution of $\g$. The right action of $\ab_-$ on
$N_+$ commutes with the right action of the Lie group $A_+ \subset N_+$ of
$\ab_+$. Therefore each element of $\ab_-$ gives rise to a vector field on
$N_+/A_+$.

In the principal gradation, the Lie algebra $\ab_-$ is linearly spanned by
elements $p_m$ of degrees $-m \in -I$. Denote by $\mu_m$ the corresponding
vector fields on $N_+/A_+$. We will show that these vector fields coincide
with the {\em hamiltonian vector fields} defined by Toda integrals.

More precisely, consider $\pi_0$ as the algebra of regular functions on an
infinite-dimensional affine space $U$ with coordinates $u_i^{(n)},
i=1,\ldots,l, n\geq 0$. We will show, cf. Theorem 2, that $U$ is isomorphic
to $N_+/A_+$ and the action of the operator $Q_i$ on $\pi_0$ coincides with
the left infinitesimal action of the $i$th generator $e_i$ of $\n_+$ on the
space of regular functions on $N_+/A_+$.

The space $U$ is equipped with a generalized Hamiltonian structure in the
sense of Gelfand-Dickey-Dorfman \cite{gd1,gd2,gdorf}. This structure allows
us to associate to any local functional $F$ in ${\bold u}$, a vector field
$\xi_F$ on the space $U$. We will say that $\xi_F$ is the hamiltonian
vector field defined by $F$. On the other hand, we show that the space of
Toda integrals is isomorphic to the dual space $\ab_+^*$ of $\ab_+$,
cf. Theorem 1 below and \cite{monte}. Thus, Toda integrals have degrees $-m
\in -I$. Denote by $\eta_m$ the vector field corresponding to the Toda
integral of degree $-m \in -I$.

Theorem 3 states that under the isomorphism $N_+/A_+ \simeq U$, we have:
$\eta_m = \mu_m$. In particular, this means that the vector fields
$\eta_m$ commute with each other. Therefore Toda integrals are in
involution with respect to the Poisson bracket $\{\cdot ,\cdot\}$. This
proves complete integrability of affine Toda equations.

\subsection{}In the course of proving this fact, we came across an interesting
subalgebra of the Lie algebra of vector fields on $N_+$. On $N_+$, we have
a right action of $\g$ and a left action of $\n_+$. Denote by $e_i^L$ the
vector field on $N_+$, which corresponds to the left action of the
generator $e_i$ of $\n_+$. Let ${\cal L}$ be the Lie algebra of vector
fields $\al$ on $N_+$, such that $[e_i^L,\al] = -F_i(\al) e_i^L$ for all
$i$, where $F_i(\al)$ is a function on $N_+$. Thus, ${\cal L}$ preserves a
certain foliation on $N_+$, and it seems that if $\g \neq \sw_2$, then
${\cal L}$ coincides with the Lie algebra of global vector fields on
$B_-\backslash G$.

The Lie algebra ${\cal L}$ can be defined for an arbitrary Kac-Moody
algebra $\g$, and one can show that $\g$ (acting from the right) is
contained in ${\cal L}$. In the Appendix we show that if $\g$ is
finite-dimensional and $\g \neq \sw_2$, then ${\cal L}=\g$, and if $\g$ is
affine, then ${\cal L}$ is the semi-direct product of $\g$ and a Lie
algebra of vector fields on the disc.

\subsection{} Let us now return to Toda integrals. Using vector fields
$\eta_m$ we can construct a hierarchy of differential equations on the
space $U$:
\begin{equation}    \label{mkdv}
\frac{\pa u_i^{(n)}({\bold t})}{\pa t_m} = \eta_m \cdot u^{(n)}_i({\bold
t}), \quad \quad m \in I; i=1,\ldots,l; n\geq 0,
\end{equation}
where ${\bold t} = (t_m)_{m \in I}$ are the times of the hierarchy. We will
show that the vector field $\xi_{H_1}$ coincides with $\pa_t$. Therefore
the equations with $m=1$ read simply as $\pa u_i^{(n)}/\pa t_1 =
\pa_t u_i^{(n)}$, and hence $t_1 \equiv t$.

It is known that Toda integrals coincide with integrals of motion of the
generalized mKdV hierarchy associated to $\g$ \cite{DS,ds,kw,w}, so that
equations \eqref{mkdv} are equations of the mKdV hierarchy. This fact can
be proved directly: using our geometric formalism we can rewrite equations
\eqref{mkdv} in the ``zero-curvature'' form and check that they
coincide with the mKdV hierarchy. In the case $\g=\su$ this follows from
results of Enriquez \cite{E}. Thus we obtain a new proof of integrability
of generalized mKdV (and therefore KdV) equations.

We can also obtain solutions of the mKdV hierarchy using the (rational)
action of the Lie group $A_-$ of $\ab_-$ on $U$. Indeed, consider the
following element of $A_-$: $$g({\bold t}) = \exp \left( \sum_{m \in I} t_m
\eta_m \right) = \exp \left( \sum_{m \in I} t_m p_m \right).$$ This
element operates on the manifold $B_- \backslash G/A_+$ from the right.

Now choose an initial condition $x({\bold 0}) \in U \subset B_-
\backslash G/A_+$ and put
\begin{equation}    \label{solution}
x({\bold t}) = x({\bold 0}) \cdot g({\bold t}).
\end{equation}
One can show that $x({\bold t})$ lies in the ``big cell'' $U \subset B_-
\backslash G/A_+$ for almost all values of the $t_m$'s (compare with
\cite{wilson1,wilson2}). But then the
$u_i$--coordinates $u_i({\bold t})$ of $x({\bold t}) \in U$ give solutions
of the mKdV hierarchy \eqref{mkdv}. It is evident that all solutions, for
which the initial function $u_i(t,0,0,\ldots)$ is smooth at least for one
value of $t$, can be obtained this way.

\subsection{} The connection between the KdV and mKdV hierarchies and
affine groups has been studied before in the framework of tau-functions,
infinite Grassmanians and dressing transformations,
cf. \cite{djkm,sw,wilson1,wilson2,cherednik}, and more recent papers
\cite{kacw,bb,hm,adberg}.

However, previously it was established through the so-called Baker
functions, of which our $u_i$'s are, roughly speaking, logarithmic
derivatives (for a more precise statement, cf. \cite{wilson1,wilson2}). The
difference with our approach is that instead of using Baker functions, we
identify the mKdV variables $u_i^{(n)}$'s directly with special coordinates
on the big cell of the flag manifold modulo $A_+$. We do that using the
nilpotent action on $U$ induced by the hamiltonian of the Toda equation.

\subsection{}
As was mentioned above, the hamiltonian form of the mKdV and Toda
hierarchies can be obtained from the generalized Hamiltonian structure on
the space $U$, which has been studied by Gelfand, Dickey and Dorfman
\cite{gd1,gd2,gdorf}. This structure can be quantized, giving the vertex
operator algebra structure on $\pi_0$. This allows us to define the space
of quantum Toda integrals.  Generally, the dimension of the space of
quantum integrals of a given degree could be smaller than the dimension of
the space of classical integrals of the same degree. However, in
\cite{ff:toda,monte} we proved that in our case these dimensions are the
same in all degrees. In other words, we proved that all classical Toda
integrals can be quantized. The quantum integrals are integrals of motion
of certain deformations of conformal field theories, which were mentioned
at the beginning of this Introduction.

It would be interesting to understand how the Gelfand-Dickey-Dorfman
structure on $U$ is related to the Lie-Poisson structure on the Borel group
of $G$ in the sense of Drinfeld \cite{drinfeld}. Quantization of
Lie-Poisson structures leads to quantum groups \cite{drinfeld1}, and
quantum groups indeed appeared in our construction of quantum integrals of
motion, cf. \cite{monte}. Roughly speaking, the action of the Lie algebra
$\n_+$ on $\pi_0$, which plays a crucial role in our classical
construction, becomes an action of the quantized universal enveloping
algebra $U_q(\n_+)$. Thus we see that quantization of the geometric
constructions given in this paper should involve both vertex operator
algebras and quantum groups.

Finally, let us remark that one can associate analogues of mKdV hierarchies
to arbitrary Heisenberg subalgebras of an affine algebra, cf.
\cite{deg,hm,adberg}. In this work we only consider the case of principal
subalgebras. Generalization of our approach to arbitrary Heisenberg
subalgebras is straightforward, but we will leave the details for a
separate publication.

\subsection{} The paper is organized as follows. In Sect. 2 we recall the
hamiltonian formalism of affine Toda field theories, following
Gelfand-Dickey-Dorfman and Kuperschmidt-Wilson. Then in Sect. 3 we show
that the Toda hamiltonian
\eqref{H} gives rise to an action of the nilpotent subalgebra $\n_+$ of
$\g$ on the algebra of differential polynomials $\pi_0$. In Sect. 4 we give
a geometric construction of modules contragradient to Verma modules over
$\g$ in the space of functions on $N_+$ and homomorphisms between them,
following \cite{kos}. We also prove an important \propref{P}. Using these
results, we prove in Sect. 5 that the algebra $\pi_0$ is isomorphic to the
algebra of regular functions on the space $N_+/A_+$. In Sect. 6 we use the
Bernstein-Gelfand-Gelfand resolution and results of Sect. 4 to show that
the space of Toda integrals is isomorphic to the first cohomology of a
complex $F^*(\g)$ and to the dual space of $\ab_+$. We explain in Sect. 7
how to attach explicitly a Toda integral to a class in the first cohomology
of $F^*(\g)$, and then obtain a crucial formula \eqref{main} for the
commutator of the vector fields $\eta_m$ and $Q_i$. Using this formula we
prove in Sect. 8 that the hamiltonian vector field $\eta_m$ defined by a
Toda integral coincides with the vector field $\mu_m$, which corresponds to
the right action of an element of the Lie algebra $\ab_-$ on $N_+/A_+$. In
the Appendix we study the subalgebra ${\cal L}$ of the Lie algebra of
vector fields on $N_+$.

\section{Hamiltonian formalism}    \label{hf}
Let $\g$ be an affine algebra, twisted or non-twisted, of rank $l+1$. To
$\g$ one canonically associates a finite-dimensional Lie algebra
$\bar{\g}$, whose Dynkin diagram is obtained from the Dynkin diagram of
$\g$ by deleting its $0$th node. We have the Cartan decomposition: $\gb =
\bar{\n}_- \oplus \hh \oplus \bar{\n}_+$, where $\hh$ is the Cartan
subalgebra and $\bar{\n}_\pm$ are the nilpotent subalgebras of $\gb$.

We consider $\g$ as defined over the formal Laurent power series $\C((t))$,
with the topology of inverse limit. The Lie algebra $\g$ is the universal
central extension of a loop algebra $\g'$ adjoined with a $\Z$-grading
operator $d$. If $\g$ is non-twisted, $\g' = \gb \otimes
\C((t))$; if $\g$ is twisted, $\g'$ consists of Laurent series, which have
special properties with respect to an automorphism of $\gb$,
cf. \cite{kac}.

Recall that the Lie algebra $\g$ has the Cartan decomposition: $\g = \n_+
\oplus \h \oplus \n_-$, where $\n_+$ and $\n_-$ are the nilpotent
subalgebras, and $\h$ is the Cartan subalgebra of $\g$. For a non-twisted
affine algebra $\g$, $\n_+ = (\bar{\n}_+ \otimes 1) \oplus (\gb \otimes
\C[[t]])$.

We have: $\h = \bar{\h} \otimes 1 \oplus \C d \oplus \C K$, $K$ being a
generator of the center. We have a linear basis $h_1,\ldots,h_l,d$ in
$\h$.  The upper and lower nilpotent Lie subalgebras $\n_+$ and $\n_-$ are
generated by elements $e_i, i=0,\ldots,l$, and $f_i, i=0,\ldots,l$, which
satisfy the Serre relations \cite{kac}: $$(\on{ad} e_i)^{-a_{ij}+1} \cdot
e_j = 0, \quad \quad (\on{ad} f_i)^{-a_{ij}+1} \cdot f_j = 0.$$

The dual space $\h^*$ to $\h$ is linearly spanned by functionals
$\al_0,\ldots,\al_l$, and $\Lambda_0$ \cite{kac}. The value of $\al_j$ on
the $i$th generator $h_i$ of the Cartan subalgebra of $\g$ is equal to the
element $a_{ij}$ of the Cartan matrix of $\g$. Denote by $(\cdot,\cdot)$
the invariant scalar product on $\h^*$, normalized as in \cite{kac}. Let
$a_i, i=0,\ldots,l$, be the labels of the Dynkin diagram of $\g$
\cite{kac}. They satisfy $$\sum_{0\leq j\leq l} a_j a_{ij} = 0, \quad \quad
\forall i=0,\ldots,l.$$

Let $U$ be the inverse limit of finite-dimensional linear spaces with
coordinates $u_i^{(n)}, 1 \leq i\leq l; 0\leq n \leq N$. Denote by $\pi_0$
the algebra of algebraic functions on $U$; this is a free polynomial
algebra with generators $u_i^{(n)}, i=1,\ldots,l, n\geq 0$. Define a
derivation $\pa$ on $\pi_0$ by putting $\pa u_i^{(n)} = u_i^{(n+1)}$, and
extending it to arbitrary elements of $\pi_0$ by the Leibnitz rule. We can
consider $\pa$ as a vector field on $U$: $$\pa = \sum_{1\leq i\leq l, n\geq
0} u_i^{(n+1)} \frac{\pa}{\pa u_i^{(n)}}.$$

We call $\pi_0$ the algebra of differential polynomials. Introduce a
$\Z$-grading on $\pi_0$ by putting $\deg u_i^{(n)} = -n-1$. With respect to
this grading, the derivative $\pa$ is a homogeneous linear operator of
degree $-1$.

We are going to introduce a hamiltonian structure on $U$ in the generalized
sense of Gelfand, Dickey and Dorfman \cite{gd1,gd2,gdorf}. We now briefly
recall their general definition of hamiltonian operator.

Let $\ab$ be a Lie algebra. A complex $\Omega = \oplus_{j\geq 0}
\Omega^j$ with a differential $d: \Omega^j \arr \Omega^{j+1}, j\geq 0$, is
called an $\ab$--complex, if for any $a \in \ab$ there is a linear map
$i_a: \Omega^j \arr \Omega^{j-1}, j>0$, such that $i_a i_b + i_b i_a = 0$
and $[i_a d + d i_a, i_b] = i_{[a,b]}$.

Given an $\ab$--complex, a linear operator $H: \Omega^1 \arr \ab$ is called
{\em hamiltonian operator} if (1) $H$ is antisymmetric: $i_{H (\omega_1)}
(\omega_2) = - i_{H (\omega_2)} (\omega_1)$ for any $\omega_1, \omega_2 \in
\Omega^1$; (2) $[H,H] = 0$, where $[\cdot,\cdot]$ is the Schoutens bracket,
cf. \cite{gdorf}.

The standard example of an $\ab$--complex is the de Rham complex of a
manifold $M$ with the Lie algebra of vector fields on $M$ as $\ab$. Then
the notion of hamiltonian operator is equivalent to the standard notion of
hamiltonian structure on $M$.

Now fix a set $Z$ of vector fields on $M$ and define $\ab_Z$ to be the
stabilizer of $Z$ in $\ab$. Denote by $\Omega_0$ the image of $L_z = i_z d
+ d i_z, z\in Z$, in $\Omega$, and put $\Omega_Z = \Omega/\Omega_0$. Then
$\Omega_Z$ is an $\ab_Z$--complex.

Such a complex naturally arises in our context. We take as $M$ the space
$U$. Then $\Omega^j$ consists of differential forms, which can be
represented as finite sums $$\omega = \sum
\omega_{i_1,\ldots,i_j}^{n_1,\ldots,n_j} du_{i_1}^{(n_1)} \wedge \ldots
\wedge d u_{i_j}^{(n_j)},$$ where
$\omega_{i_1,\ldots,i_j}^{n_1,\ldots,n_j}$ is a polynomial in
$u_i^{(n)}$'s. In particular, $\Omega^0 = \pi_0$. The Lie algebra $\ab$
consists of all vector fields $$\sum_{n\geq 0} \sum_{1\leq i\leq l} X_{i,n}
\frac{\pa}{\pa u_i^{(n)}},$$ where $X_{i,n} \in \pi_0$ and the sum may be
infinite. The map $i_a$ is defined as the usual contraction of a
differential form with vector field $a$.

Now put $Z = \{ \pa \}$. The Lie algebra $\ab_Z$ consists of all vector
fields, which commute with $\pa$. We have: $$[\pa,\pa_i^{(n)}] =
-\pa_i^{(n-1)}.$$ Therefore such vector fields have the form $$\sum_{ n\geq
0} \sum_{1\leq i\leq l} (\pa^n X_i) \frac{\pa}{\pa u_i^{(n)}}$$ for some
$X_i \in \pi_0, i=1,\ldots,l$.

The space $\Omega^0_\pa$ is the quotient of $\Omega^0 = \pi_0$ by the
action of $\pa$. Let $\F_0$ be the quotient of $\Omega^0_\pa$ by the
constants. We can introduce a $\Z$-grading on $\F_0$ by adding $1$ to the
grading induced from $\pi_0$. The space $\F_0$ can be interpreted as the
space of local functionals of the form \eqref{lf}, since the integral of a
total derivative or a constant is equal to $0$. Denote by $\int$ the
projection $\pi_0 \arr \F_0$.

The space $\Omega^1_\pa$ is the quotient of $\Omega^1$ by the action
of $\pa$. Therefore any element of this space can be uniquely represented
in the form $$\sum_{1\leq i\leq l} Y_i d u_i^{(0)}.$$ The differential $d:
\Omega^0_\pa \arr \Omega^1_\pa$ is given by the formula $$d \bar{P} =
\sum_{1\leq i\leq l} \frac{\delta P}{\delta u_i} d u_i^{(0)} \, \, \on{mod}
\, \on{Im} \pa,$$ where $\bar{P}$ is the projection of $P \in \pi_0$ on
$\Omega^0_\pa$ and $$\frac{\delta P}{\delta u_i} = \sum_{n\geq 0}
\sum_{1\leq i\leq l} (-\pa)^n \frac{\pa P}{\pa u_i^{(n)}}$$ is the
variational derivative with respect to $u_i$. This map is well-defined,
because the variational derivative of a total derivative is $0$.

Now consider the operator $H: \Omega^1_\pa \arr \ab_\pa$, which sends
$$\sum_{1\leq i\leq l} Y_i d u_i^{(0)} \, \, \on{mod}
\, \on{Im} \pa$$ to $$\sum_{ n\geq
0} \sum_{1\leq i,j\leq l} (\al_i,\al_j) (\pa^{n+1} Y_i) \frac{\pa}{\pa
u_j^{(n)}}.$$ This operator is hamiltonian, cf. \cite{gdorf}. Using this
operator we can now assign to any element $P \in \pi_0$ a vector field
$H(d\bar{P})$ on $U$. Denote the corresponding derivation of $\pi_0$ by
$\xi_P$. Since constants are annihilated by $d$, $\xi_P$ depends only on
the image of $P$ in $\F_0$, $\int P$. Sometimes we will write $\xi_{\int
P}$ instead of $\xi_P$. We easily find that
\begin{equation}    \label{vf}
\xi_P = \sum_{1\leq i\leq l, n\geq 0} ( \pa^{n+1} \cdot \delta_i P)
\fp{i}{n},
\end{equation}
where we put $$\delta_i P = \sum_{n\geq 0} \sum_{1\leq j\leq l}
(\al_i,\al_j) (-\pa)^n \frac{\pa P}{\pa u_j^{(n)}}, \quad \quad
i=1,\ldots,l.$$ In particular,
\begin{equation}    \label{pa}
\pa = \sum_{1\leq i\leq l, n\geq 0} u_i^{(n+1)}
\frac{\pa}{\pa u_i^{(n)}} = \xi_P, \quad \quad P = \frac{1}{2} \sum_{1\leq
i\leq l} u_i^{(0)} u^{(0)i},
\end{equation}
where $u^{(0)i}, i=1\ldots,l,$ are vectors dual to $u_i^{(0)},
i=1\ldots,l,$ with respect to the scalar product defined by
$(\cdot,\cdot)$.

It is proved in \cite{gdorf} that if $H$ is a hamiltonian operator, then
one can define a structure of Lie algebra on $\Omega^0$ by the formula $$\{
F_1,F_2 \} = H(dF_1) \cdot F_2.$$ Moreover, the map $H \circ d: \Omega^0
\arr \ab$ becomes a homomorphism of Lie algebras. If $\Omega$ is the de Rham
complex on a manifold, this is of course the standard definition of Poisson
bracket.

In our case we obtain a Lie bracket on $\Omega^0_\pa$ and hence on $\F_0$
by putting
\begin{equation}    \label{pb}
\{\int P,\int R\} = \int (\xi_P \cdot R).
\end{equation}

\begin{rem}
If we view elements of $\F_0$ as functionals on the space of
functions on the circle with values in the Cartan subalgebra of $\bar{\g}$,
${\bold u}(t) = (u_1(t),\ldots,u_l(t))$, then we can interpret formula
\eqref{pb} as a Poisson bracket between such functionals, cf.
\cite{gd1,gd2,monte}.

We remark that conventions in this paper differ slightly from those in
\cite{monte}.\qed
\end{rem}

Now we extend the map $\xi$ from $\pi_0$ to a larger space, following
Kuperschmidt and Wilson \cite{kw,w}.

Denote by $\La$ the root lattice in $\h^*$, which is spanned by
$\al_0,\ldots,\al_l$.

Let us formally introduce variables $\phi_i, i = 0,\ldots,l$. For any
element $\la = \sum_{0\leq i\leq l} \la_i \al_i$ of $\La$, define the linear
space $\pi_\la = \pi_0 \otimes e^{\laa}$, where $\bar{\la} = \sum_{0\leq
i\leq l}
\la_i \phi_i$, equipped with an action of $\pa$ by the formula
\begin{equation}    \label{pala}
\pa \cdot (P \otimes e^{\laa}) = (\pa P) \otimes e^{\laa} + \left(
\sum_{0\leq i\leq l} \la_i u_i^{(0)} P \right) \otimes e^{\laa},
\end{equation}
where we put $$u_0^{(n)} = -\frac{1}{a_0} \sum_{1\leq i\leq l} a_i
u_i^{(n)}.$$ This formula means that we put $\pa \phi_i = u_i^{(0)}$.

Introduce a $\Z$--grading on $\pi_\la$ by putting $\deg e^{\laa} =
(\rho^\vee,\la)$, where $\rho^\vee \in \h^*$ is such that
$(\rho^\vee,\al_i) = 1, i=0,\ldots,l$.

Let $\F_\la$ be the quotient of $\pi_\la$ by the subspace of total
derivatives and $\int$ be the projection $\pi_\la \arr \F_\la$. We define a
$\Z$--grading on $\F_\la$ by adding $1$ to the grading induced from
$\pi_\la$.

For any $P \in \F_0$ the derivation $\xi_P: \pi_0 \arr \pi_0$ can be
extended to a linear operator on $\oplus_{\la \in \La} \pi_\la$ by the
formula $$\xi_P = \sum_{1\leq i\leq l, n\geq 0} ( \pa^{n+1} \cdot \delta_i
P) \frac{\pa}{\pa u_i^{(n)}} + \sum_{1\leq i\leq l} \delta_i P
\frac{\pa}{\pa \phi_i},$$ where $\pa/\pa
\phi_i \cdot (S e^{\laa}) = \la_i S e^{\laa}$. This defines a structure
of $\F_0$-module on $\pi_\la$.

For any $P \in \pi_0$ the operator $\xi_P$ commutes with the action of
derivative. Hence we obtain the structure of an $\F_0$-module on $\F_\la$
that is a map $\{ \cdot,\cdot \}: \F_0 \times \F_\la \arr
\F_\la$: $$\{ \int P,\int R \} = \int \xi_{P} \cdot R.$$

Similarly, any element $R \in \pi_\la$ defines a linear operator $\xi_R$,
acting from $\pi_0$ to $\pi_\la$ and commuting with $\pa$:
\begin{equation}    \label{oper}
\xi_{S e^{\laa}} = \sum_{1\leq i\leq l, n\geq 0} \pa^n \left( \pa
(\delta_i S \cdot e^{\laa}) -  S \, \frac{\pa e^{\laa}}{\pa
\phi_i} \right) \fp{i}{n}.
\end{equation}
The operator $\xi_R$ depends only on the image of $R$ in
$\F_\la$. Therefore it gives rise to a map $\{ \cdot,\cdot \}: \F_\la \times
\F_0 \arr \F_\la$. We have for any $P \in \F_0, R \in
\F_\la$: $$\int \xi_R \cdot P = - \int \xi_P \cdot R.$$ Therefore our
bracket $\{ \cdot,\cdot \}$ is antisymmetric.

We also have for any $P \in \F_0, R \in \oplus_{\la \in \La}
\F_\la$:
\begin{equation}    \label{com}
\xi_{\{ P , R\}} = [\xi_P , \xi_R].
\end{equation}
This can be proved in the same way as in the case when $R \in \F_0$.

\section{Toda integrals and nilpotent action}    \label{nilp}
Denote by $\q_i, i=0,\ldots,l,$ the linear operators $-\xi_{\int
e^{-\phi_i}}: \pi_0 \arr \pi_{-\al_i}$. We have:
\begin{equation}    \label{borel}
\q_i = - \sum_{n\geq 0} (\pa^n e^{-\phi_i}) \pa_i^{(n)},
\end{equation}
where $$\pa_i^{(n)} = \sum_{1\leq j\leq l} (\al_i,\al_j) \frac{\pa}{\pa
u_j^{(n)}}.$$ Clearly, $\pa^n e^{-\phi_i} = B_i^{(n)} e^{-\phi_i},$ where
$B_i^{(n)}$'s are certain polynomials in $u_i^{(m)}$'s. These polynomials,
which are closely related to Schur's polynomials, are called Fa\`a di Bruno
polynomials. More precisely, Fa\`a di Bruno polynomials are $B_i^{(n)}$
with $u_i^{(m)}$'s replaced by $-u_i^{(m)}$'s. These polynomials appear
ubiquitously in the theory of solitons, cf., e.g., \cite{dickey}. They
satisfy a recurrence relation:
\begin{equation}    \label{rec}
B_i^{(n)} = -u_i^{(0)} B_i^{(n-1)} + \pa B_i^{(n-1)},
\end{equation}
with the initial condition $B_i^{(0)} = 1$.

We can now define derivations $Q_0,\ldots,Q_l$ of $\pi_0$ by the
formula
\begin{equation}    \label{qi}
Q_i = - \sum_{n\geq 0} e^{\phi_i} (\pa^n e^{-\phi_i}) \pa_i^{(n)} = -
\sum_{n\geq 0} B_i^{(n)} \pa_i^{(n)}.
\end{equation}

The following statement was proved in \cite{monte}, Proposition 2.2.8. It
also follows from \thmref{iso} below.

\begin{prop}    \label{serre}
The operators $Q_i$ satisfy the Serre relations of the Lie
algebra $\g$: $$(\mbox{ad} \, Q_i)^{-a_{ij}+1} \cdot
Q_j = 0,$$ where $\|a_{ij}\|$ is the Cartan matrix of $\g$.
\end{prop}

According to \propref{serre}, we obtain a structure of $\n_+$-module on
$\pi_0$ by assigning to each generator $e_i$ of the nilpotent Lie
subalgebra $\n_+$ of $\g$ the operator $Q_i: \pi_0 \arr \pi_0,
i=0,\ldots,l$.

Each operator $\q_i$ commutes with the action of $\pa$ and induces a linear
operator $\Q_i = -\{\int e^{-\phi_i},\cdot\}: \F_0 \arr \F_{-\al_i}$. Thus,
up to a sign, the sum of the operators $\Q_i$ coincides with the operator of
the bracket $$\{ H,\cdot \}: \F_0 \arr \oplus_{0\leq i\leq l}
\F_{-\al_i}$$ with the hamiltonian \eqref{H} of the Toda field theory
associated to $\g$. The elements of the space of local functionals, $\F_0$,
which lie in the kernel of this operator, are called local integrals of
motion of the corresponding Toda theory. They are conserved with respect to
the Toda equation. For brevity, we will call them {\em Toda integrals}.

Thus, by definition, the space of Toda integrals is the intersection of
kernels of the operators $\Q_i: \F_0 \arr \F_{- \al_i}, i=0,\ldots,l$.

Denote by $I$ the set of exponents of $\g$ modulo the Coxeter number. In
this paper we will give a new proof of the following result.

\begin{thm}    \label{span}
{\em The space of Toda integrals is linearly spanned by elements $H_m, m
\in I$, where $\deg H_m = -m$.}
\end{thm}

\thmref{span} will be proved in \secref{prone}. We will show that the space
of Toda integrals is isomorphic to the first cohomology of a certain
complex $F^*(\g)$ constructed from the dual of the
Bernstein-Gelfand-Gelfand (BGG) resolution of $\g$. In order to do that we
will realize modules contragradient to Verma modules over $\g$ in the space
of functions on the nilpotent group $N_+$, cf. \secref{geocon}, and
establish an isomorphism between the space $\pi_0$ and the space of
functions on $N_+/A_+$, cf. \thmref{iso}.

\begin{rem} The map $\xi: \pi_0 \arr \on{End} \pi_0$ can be quantized in
the following sense. Put $\pi_\la^\hbar = \pi_\la \otimes
\C[[\hbar]]$. There exists a map $\xi^\hbar:
\pi_0^\hbar \arr \on{End} \pi_0^\hbar$ such that (1) $\xi^\hbar$ factors
through $\F_0^\hbar = \F_0 \otimes \C[[\hbar]] = \pi_0^\hbar/\pa \cdot
\pi_0^\hbar$; (2) the bracket $\{ \cdot,\cdot \}_\hbar: \F_0^\hbar \times
\F_0^\hbar \arr \F_0^\hbar$, defined by the formula $$\{ \int P,\int R \} =
\int \xi_P \cdot R,$$ where $\int$ denotes the projection $\pi_0^\hbar \arr
\F_0^\hbar$, is a Lie bracket; (3) $\xi^\hbar = \hbar \xi^{(1)} +
\hbar^2(\ldots)$, and the map $\pi_0 \arr
\on{End} \pi_0$ induced by $\xi^{(1)}$ coincides with $\xi$.

Such a map $\xi^\hbar$ can be defined using the vertex operator algebra
structure on $\pi_0$. This is explained in \cite{monte}, Sect. 4 (where
$\hbar$ is denoted by $\beta^2$). Here we will give an explicit formula for
$\xi^\hbar: \pi_0 \arr \on{End} \pi_0$.

Introduce a Heisenberg Lie algebra with generators $b_i(n), i=1,\ldots,l; n
\in \Z$, and ${\bold 1}$, and relations $$[b_i(n),b_j(m)] = \hbar \cdot n
\cdot (\al_i,\al_j) \delta_{n,-m} {\bold 1}, \quad \quad [b_i(n),{\bold 1}]
= 0.$$ We can make $\pi_0$ into a module over this Lie algebra by defining
the following action of the generators: $$b_i(n) =
\frac{u_i^{(-n-1)}}{(-n-1)!}, \quad \quad n<0,$$ $$b_i(n) = \sum_{1\leq j
\leq l} \hbar \cdot n! \cdot (\al_i,\al_j) \frac{\pa}{\pa u_j^{(n-1)}}, \quad
\quad n>0,$$ and $$b_i(0) = 0, \quad \quad {\bold 1} = \on{Id}.$$

Denote $$b_i(z) = \sum_{n\in \Z} b_i(n) z^{-n-1}.$$ For a monomial
$u_{i_1}^{(n_1)} \ldots u_{i_m}^{(n_m)} \in \pi_0$ define the following
formal power series $$:\pa_z^{-n_1} b(z) \ldots \pa_z^{-n_m} b(z):,$$ where
the dots stand for the normal ordering \cite{monte}. The Fourier
coefficients of this series are linear operators acting on
$\pi_0^\hbar$. By linearity we obtain a map $Y(\cdot,z): \pi_0^\hbar \arr
\on{End}
\pi_0^\hbar[[z,z^{-1}]]$. This map defines a structure of vertex operator
algebra on $\pi_0$, which depends on $\hbar$ \cite{monte}. For $P \in
\pi_0^\hbar$ denote by $\xi_P^\hbar$ the linear endomorphism of
$\pi_0^\hbar$ given by the residue, i.e. the $(-1)$st Fourier component, of
$Y(P,z)$. This gives us a map $\xi^\hbar: \pi_0^\hbar \arr \on{End}
\pi_0^\hbar$, which satisfies the conditions above.

Thus, the Gelfand-Dickey-Dorfman structure on $\pi_0$ can be viewed as a
classical limit of the structure of vertex operator algebra on
$\pi_0$. Note that one can also define classical limits of other Fourier
components of $Y(P,z)$ for any $P \in \pi_0$. Altogether they satisfy
certain properties, which are corollaries of axioms of vertex operator
algebra.

In \cite{monte} we also defined quantum deformations of the maps $\xi:
\pi_0 \arr \on{End} \pi_\la$ and $\pi_\la \arr \on{Hom}(\pi_0,\pi_\la)$. This
enabled us to quantize the operators $\q_i: \pi_0 \arr \pi_{-\al_i}$ and
$\Q_i: \F_0 \arr \F_{-\al_i}$ and define the space of quantum Toda
integrals as the intersection of kernels of the quantum operators
$\Q_0,\ldots,\Q_l$. For these quantum operators one has an analogue of
\propref{serre}: in a certain sense they satisfy the Serre relations of the
quantized universal enveloping algebra $U_q(\n_+)$, where $q=\exp(\pi i
\hbar)$, $\hbar$ being the deformation parameter. This important
observation goes back to \cite{bmp}. Using this fact, we proved in
\cite{ff:toda,monte} that all classical Toda integrals can be quantized (at
least in the case when all exponents of $\g$ are odd and the Coxeter number
is even).\qed
\end{rem}

\section{Geometric construction}    \label{geocon}
Let $G$ be the Lie group of $\g$. This group is the universal central
extension of the loop group of $\bar{G}$ or its subgroup, if $\g$ is
twisted. The group $G$ is also defined over formal Laurent series, and it
is also equipped with the topology of inverse limit. We consider the
induced topology on Lie subgroups of $G$.

Let $B_+$ and $B_-$ be the Borel subgroups of $G$. Their projections on the
loop group of $\bar{G}$ consist of its $\C[[t]]$-points (respectively,
$\C[t^{-1}]$-points), whose image in the constant Lie subgroup $\bar{G}$ of
$G$ belongs to the finite-dimensional Borel subgroup $\bar{B}_+$
(respectively, $\bar{B}_-$), or its invariant part with respect to an
automorphism, if $\g$ is twisted. Let $H = B_+ \cap B_-$ be the Cartan
subgroup of $G$.

Let $N_+$ be radical of $B_+$; this is the Lie group of $\n_+$. The group
$N_+$ is a prounipotent proalgebraic group, which is the inverse limit of
finite-dimensional algebraic groups $N_+^{(n)}\backslash N_+, n>0$,
isomorphic to finite-dimensional affine spaces. Here $N_+^{(n)}$ denotes
the subgroup of $N_+$, which consists of elements, which are equal to $1$
modulo $t^n$.  The exponential map $\exp: \n_+ \arr N_+$ is an
isomorphism. By the space of functions on $N_+$ we will always mean the
space of algebraic functions, which is the inductive limit of the spaces of
algebraic functions on $N_+^{(n)}\backslash N_+, n>0$.

Consider $p = \sum_{0\leq i\leq l} a_i e_i$, a principal element of $\n_+$,
where $e_i, i=0,\ldots,l$, are the generators of $\n_+$, and $a_i,
i=0,\ldots,l$, are the labels of the Dynkin diagram of $\g$. Denote by
$\ab_+$ the {\em principal commutative Lie subalgebra} of $\n_+$, which is
the centralizer of $p$ in $\n_+$.

The Lie algebra $\g$ is graded by weights of the Cartan subalgebra $\h$. We
can also introduce the {\em principal} $\Z$--grading on $\g$ by putting
$\deg h_i = \deg d = \deg K = 0, \deg e_i = -\deg f_i = 1,
i=0,\ldots,l$. It is known that with respect to this grading the Lie
algebra $\ab_+$ is linearly spanned by elements of degrees equal to the
exponents of $\g$ modulo the Coxeter number. It is also known that the Lie
algebra $\n_+$ splits into the direct sum $\n_+ = \on{Ker} p
\oplus \on{Im} p$, where $\on{Ker} p = \ab_+$ and $\on{Im} p = \oplus_{j>0}
\n_+^j$ is the direct sum of homogeneous components $\n_+^j$ of degree
$j\geq 0$, each having the same dimension $l$. For a proof of these facts,
cf. \cite{Kac}, Proposition 3.8 (b).

The Lie subalgebra $\ab_+$ is the positive half of the principal Heisenberg
subalgebra $\widehat{\ab}$ of $\g$. This Heisenberg algebra is the central
extension by $\C K$ of the commutative subalgebra $\ab$ of the loop algebra
$\g' = [\g,\g]/\C K$, which is the centralizer of $p$ in $\g'$. We
have $\ab = \ab_+ \oplus \ab_-$, where $\ab_-$ is a commutative Lie
subalgebra of $\n_-$. The Lie algebra $\ab_-$ is linearly spanned by
generators of degrees equal to minus the exponents of $\g$ modulo the
Coxeter number. Let us choose for each $m
\in I$ a linear generator $p_m$ of $\ab_-$ degree $-m$. We choose
\begin{equation}    \label{p1}
p_1 = \sum_{0\leq i\leq l} \frac{(\al_i,\al_i)}{2} f_i.
\end{equation}

The group $N_+$ is isomorphic to the big cell $X$ of the flag manifold $F=
\fl$, which is the orbit of the image of $1 \in G$ under the action of
$N_+$. Flag manifolds of $G$ have been studied, e.g., in
\cite{kl,kp1,kp2,ps,ka}.

The group $G$ acts on the flag manifold from the right: $(g,f) \arr f
g^{-1}, g \in G, f \in F$. This induces an infinitesimal action of $\g$ on
the big cell $X \simeq N_+$ by vector fields. The Lie algebra of vector
fields on the circle $\Ve = \C((t^k))t\pa_t$, where $k$ is the order of the
associated automorphism of $\bar{\g}$, infinitesimally acts on the group
$G$, cf., e.g., \cite{ps}. Its Lie subalgebra $\Ve_- = \C[t^{-k}] t\pa_t$
preserves the group $B_-$ and hence it maps to vector fields on $F$ and on
$X \simeq N_+$. Thus, the Lie algebra $\g \times \Ve_-$ maps to the Lie
algebra $\V$ of vector fields on $N_+$. Note that the central element $K$
maps to $0$, and the image of $d$ coincides with the image of $t \pa_t \in
\Ve_-$. Therefore this map factors through the Lie algebra $\G = \g' \times
\Ve_-$. Denote the vector field corresponding to an element $\al \in \G$ by
$\al^R$.

The Lie group $N_+$ acts on itself from the left: $(n_1,n_2) \arr n_1 n_2$
and from the right: $(n_1,n_2) \arr n_2 n_1^{-1}$. These actions induce two
infinitesimal actions of the Lie algebra $\n_+$ on $N_+$, and hence two
embeddings of $\n_+$ into the Lie algebra $\V$. Denote by $\n_+^L$ and
$\n_+^R$ the images of these embeddings. For $\beta \in \n_+$ we denote by
$\beta^L$ and $\beta^R$ the images of $\beta$ in $\n_+^L$ and $\n_+^R$,
respectively. The Lie algebra $\n_+^R$ coincides with the image of $\n_+
\subset \G$ in $\V$. Note that $\n_+^L$ and $\n_+^R$ commute with each
other.

We will now describe a geometric construction of modules contragradient to
the Verma modules over $\g$ and homomorphisms between them. This
construction is an affine analogue of Kostant's construction \cite{kos}
(cf. also \cite{z,kv,bmp}) in the case of simple Lie algebras, though our
argument is somewhat different from his. Our treatment can be easily
generalized to arbitrary Kac-Moody algebras.

For $\la \in \h^*$, denote by $\C_\la$ the one-dimensional representation
of $\goth{b}_+$, on which $\h \subset \goth{b}_+$ acts according to its
character $\la$, and $\n_+ \subset \goth{b}_+$ acts trivially. Let $M_\la$
be the Verma module over $\g$ of highest weight $\la$: $$M_\la = U(\g)
\otimes_{U(\goth{b}_+)} \C_\la.$$

Denote by $\lef\cdot,\cdot\ri$ the pairing $M_\la^* \times M_\la
\arr \C$. Let $\omega$ be the Cartan anti-involution on $\g$, which maps
generators $e_0,\ldots,e_l$ to $f_0,\ldots,f_l$ and vice versa and
preserves $\h$ \cite{kac}. It extends to an anti-involution of $U(\g)$. Let
$M^*_\la$ be the module contragradient to $M_\la$. As a linear space,
$M_\la^*$ is the restricted dual of $M_\la$. The action of $x \in \g$ on $y
\in M_\la^*$ is defined as follows: $$\lef x\cdot y,z \ri = \lef
y,\omega(x)\cdot z \ri, \quad \quad z \in M_\la.$$

Suppose first that the highest weight $\la$ is integral. Then the module
$\C_\la$ over $\goth{b}_-$ can be integrated to a module over $B_-$. To
this module we associate in the standard way an equivariant line bundle on
$F$. The module $M_\la^*$ can be realized as the space of sections of the
restriction of this line bundle to the big cell $X$. But this line bundle
can be trivialized over $X$. Therefore the module $M^*_\la$ is isomorphic
to the space of algebraic functions on $X$ with respect to the twisted
action of $\g$ by first order differential operators. For an element
$\beta$ of $\g$ this differential operator is equal to $\beta^R +
F_\la(\beta)$, where $F_\la(\beta)$ is an algebraic function on $X$. Note
that if $\beta$ is homogeneous, $F_\la(\beta)$ is also homogeneous of the
same weight.

We can interpret the functions $F_\la(\beta)$ as elements of the group
$H^1(\g,\C[X])$, where $\C[X]$ is the space of algebraic functions on $X$
considered as a $\g$-module with respect to the right infinitesimal
action. Indeed, in the standard complex of Lie algebra cohomology an
element of $H^1(\g,\C[X])$ is realized as a linear functional $f$ on $\g$
with values in $\C[X]$. Such an element defines a deformation of the
$\g$-module $\C[X]$: the deformed action of $\beta \in \g$ is obtained by
adding to the old action the operator of multiplication by $f(\beta)$.

Note that as a $\g$--module, $\C[X] = M_0^*$ is coinduced from the trivial
representation of $\goth{b}_-$. By Shapiro's lemma (cf., e.g.,
\cite{fuchs}, Sect. 1.5.4, \cite{gui}, Sect. II.7), $H^1(\g,\C[X]) \simeq
H^1(\goth{b}_-,\C) \simeq (\goth{b}_-/[\goth{b}_-,\goth{b}_-])^* =
\h^*$. We see that all elements of $H^1(\g,\C[X])$ have weight $0$. On the
other hand, functions on $X$ can only have negative or $0$ weights and the
only functions, which have weight $0$ are constants, which are invariant
with respect to the action of $\g$. Therefore the coboundary of any element
of the $0$th group of the complex, $\C[X]$, has a non-zero weight. Hence
any cohomology class from $H^1(\g,\C[X])$ canonically defines a one-cocycle
$f$, that is a map $\g \arr \C[X]$. Thus, having identified the
space of deformations with $\h^*$, we can assign to each $\la \in \h^*$ and
each $\beta \in \g$, a function on $X$ -- this is our $F_\la(\beta)$.

By linearity, if $\la = \sum_{i=0}^l \la_i \al_i$, then $F_\la(\beta) =
\sum_{i=0}^l \la_i F_i(\beta)$, where we put $F_i(\beta) =
F_{\al_i}(\beta)$. Thus, this construction works for arbitrary, not only
integral, values of $\la$.  In particular, we have: $F_\la(\beta) = 0$ for
any $\la$, if $\beta \in \n_+$, and $F_\la(\beta) = \la(\beta)$, if $\beta
\in \h$.

Thus, for any weight $\la \in \h^*$, we realize the module $M^*_\la$ as the
space of functions on $X$ equipped with the twisted action of $\g$ by first
order differential operators.

Vector $y$ from the Verma module $M_\la$ is called {\em singular vector} of
weight $\mu \in \h^*$, if $\n_+ \cdot y = 0$ and $x \cdot y = \mu(x) y$ for
any $x \in \h$. We have $M_\la \simeq U(\n_-) \cdot v_\la$, where $v_\la$,
which is called the highest weight vector, is a generator of the space
$\C_\la$.  This vector is a singular vector of weight $\la$. Any singular
vector of $M_\la$ of weight $\mu$ can be uniquely represented as $P \cdot
v_\la$ for some element $P \in U(\n_-)$ of weight $\mu - \la$. This
singular vector canonically defines a homomorphism of $\g$-modules $i_P:
M_\mu \arr M_\la$, which sends $u \cdot v_\mu$ to $(uP) \cdot v_\la$ for
any $u \in U(\n_-)$.  Denote by $i_P^*$ the dual homomorphism $M^*_\la \arr
M^*_\mu$.

There is an isomorphism $U(\n_-) \arr U(\n_+)$, which maps the generators
$f_0,\ldots,f_l$ to $-e_0,\ldots,-e_l$. Denote by $\bar{P}$ the image of $P
\in U(\n_-)$ under this isomorphism.

The homomorphism $\n_+ \arr \V$, mapping $\al \in \n_+$ to $\al^L$, can be
extended in a unique way to a homomorphism from $U(\n_+)$ to the algebra of
differential operators on $X$. Denote the image of $u \in U(\n_+)$ under
this homomorphism by $u^L$.

It turns out that using the left infinitesimal action of $\n_+$ on $X$, one
can realize the dual homomorphisms $i^*_P: M^*_\la \arr M^*_\mu$ via
differential operators on $X$.

\begin{prop}    \label{hom}
If $P \cdot v_\la$ is a singular vector in $M_\la$ of weight $\mu$, then
the homomorphism $i^*_P: M^*_\la \arr M^*_\mu$ is given by the differential
operator $\bar{P}^L$.
\end{prop}

\begin{pf} As an $\n_-$--module, the module $M_\la$ is isomorphic to
$U(\n_-)$. Hence in addition to the left action of $\g$, we have a right
action of $\n_-$ on $M_\la$: $(n,u \cdot v_\la) \arr -(u n) \cdot v_\la, n
\in \n_-, u \in U(\n_-)$. If we realize $M_\la^*$ as the space of functions
on $X$, this right action corresponds to the action of $\n_+^L$ on
$\C[X]$.

More precisely, $$\lef \beta^L \cdot x,u \cdot v_\la \ri = -
\lef x,(u\omega(\beta)) \cdot v_\la \ri, \quad \quad \forall \beta \in
\n_+, u \in U(\n_-).$$ Therefore $$\lef
\bar{P}^L \cdot x,u \cdot v_\la \ri = \lef x,(uP) \cdot v_\la \ri.$$
On the other hand, by definition, $$\lef i^*_P \cdot x,u \cdot v_\la \ri
= \lef x,i_P \cdot (u \cdot v_\la) \ri = \lef x,(uP) \cdot v_\la \ri,$$ and
Proposition follows.
\end{pf}

We can now derive the following important result.

\begin{prop}    \label{P}
\begin{itemize}
\item[(a)]
If $\al \in \V$ is such that for any $\beta \in \n_+^L$ (respectively,
$\beta \in \n_+^R$) $[\al,\beta]=0$, then $\al \in \n_+^R$ (respectively,
$\al \in \n_+^L$).
\item[(b)]
For any $\beta \in \G$ we have:
\begin{equation}    \label{key}
[e_i^L,\beta^R] = - F_i(\beta) e_i^L, \quad \quad i=0,\ldots,l.
\end{equation}
\end{itemize}
\end{prop}

\begin{pf} Part (a) is clear, because such a vector field is $\n_+^L$-- or
$\n_+^R$--invariant, and hence is uniquely defined by its value at the origin
of the group $N_+$.

To prove part (b), consider the particular case of \propref{hom}, when
$\la=0$ and $\mu=-\al_i$. It is known that vector $f_i \cdot v_0$ is a
singular vector of $M_0$ of weight $-\al_i$. By \propref{hom}, $-e_i^L$ is
a $\g$-homomorphism from $M^*_0$ to $M^*_{-\al_i}$. It is therefore a
$\G$-homomorphism, since the action of $\Ve_-$ on any $M^*_\la$ (and any
module from the category ${\cal O}$ of $\g$-modules) can be expressed in
terms of the action of $\g$ via the Sugawara construction.

Thus the operator $e_i^L$ intertwines the actions of $\G$ on $M^*_0$ and
$M^*_{-\al_i}$. But $\beta \in \G$ acts on $M^*_0$ as $\beta^R$, and on
$M^*_{-\al_i}$ as $\beta^R - F_i(\beta)$. This gives us formula
\eqref{key}.
\end{pf}

It is clear that the Cartan subalgebra $\h$ of $\g$ acts diagonally on
$\V$, and thus defines a grading of $\V$ by weights of $\g$. In particular,
vector fields $e_i^L$ and $e_i^R$ both have weight $\al_i$. Therefore we
have $[e_i^L,h_j^R] = -a_{ji} e_i^L$, and so $F_i(h_j) = a_{ji}$.

Now consider $F_i(f_j)$. Let us apply $e_k^R$ to the left and right hand
sides of formula \eqref{key} for $\beta = f_j$. Since $[e_k^R,e_i^L]=0$ by
\propref{P} (a), we obtain:
$$[e_k^R,[e_i^L,f_j^R]] = [[e_k^R,e_i^L],f_j^R] + [e_i^L,[e_k^R,f_j^R]] =
\delta_{k,j} [e_i^L,h_j^R] = -a_{ji} \delta_{k,j} e_i^L$$ $$ = - (e_k^R
\cdot F_i(f_j)) e_i^L.$$ Therefore $F_i(f_j)$ satisfies: $$(e_k^R \cdot
F_i(f_j)) = a_{ji} \delta_{k,j}.$$ There are unique functions $x_i,
i=0,\ldots,l$, on $N_+$, which have the property
\begin{equation}    \label{property}
e_k^R \cdot x_j = -e_k^L \cdot x_j = \delta_{k,j}.
\end{equation}
We see that $F_i(f_j) = a_{ji} x_j.$

\section{Isomorphism between $\pi_0$ and $\C[N_+/A_+]$}
Now denote by $A_+$ the image of $\ab_+$ in $N_+$ under the exponential
map. Let $Y$ be the homogeneous space $N_+/A_+$. On $Y$ we have the left
infinitesimal action of $\n_+$. Since the right infinitesimal action of the
Lie algebra $\ab_-$ commutes with the right action of $A_+$ on $N_+$, we
obtain a homomorphism from $\ab_-$ to the Lie algebra of vector fields on
the space $Y$.

There exists $\De \in \h$, such that $[\De,e_i] = 1$ for
$i=0,\ldots,l$. The action of $\Delta$ on $\g$ coincides with the action of
the principal $\Z$-grading. Therefore it preserves the Lie algebra $\ab_+$,
and hence $\De^R$ defines a vector field on $Y$, for which we will use the
same notation. This gives us a $\Z$-grading on the space of functions on
$Y$.

Let $y$ be an element of $\ab_-$ of degree $m$. Then the functions $F_i(y)$
are $\ab_+^R$-invariant algebraic functions on $N_+$ of degree $m$. Indeed,
the action of $y$ on $M_{\al_i}^*$ is given by $y^R+F_i(y)$ and the action
of $x \in \ab_+^R \subset \n_+^R$ is given by $x^R$. These actions must
commute (note that the central element maps to $0$ in $\V$). But since
$[y^R,x^R] = 0$, we obtain $x^R \cdot F_i(y)=0$. In particular, $F_i(p_1)$
are $\ab_+^R$--invariant algebraic functions on $N_+$ of degree $-1$.

Denote by $u_i$ the algebraic function on $Y = N_+/A_+$, corresponding to
$F_i(p_1), i=1,\ldots,l$. Formula \eqref{p1} gives:
\begin{equation}    \label{nov}
u_i = \sum_{0\leq j\leq l} (\al_i,\al_j) x_j,
\end{equation}
because $a_{ji} = 2(\al_i,\al_j)/(\al_j,\al_j)$. Let $u_i^{(m)},
i=1,\ldots,l, m\geq 0$, be the algebraic function $p_1^m \cdot
u_i^{(0)}$. Denote, as in the previous section, by $U^{(m)}$ the linear
space with coordinates $u_i^{(j)}, j=1,\ldots,l, j=0,\ldots,m$, and by $U$
the inverse limit of the spaces $U^{(m)}$.

\begin{prop}    \label{homom}
The homogeneous space $Y = N_+/A_+$ is isomorphic to the space $U$.
\end{prop}

\begin{pf}
Consider the values of the differentials $d u^{(n)}_{i,\bar{1}}$ of the
functions $u_i^{(n)}$ at the image $\bar{1} \in Y$ of $1 \in N_+$. They are
vectors in the cotangent space to $Y$ at $\bar{1}$, which is naturally
isomorphic to $(\n_+ / \ab_+)^*$. The covectors $d u^{(0)}_{i,\bar{1}},
i=1,\ldots,l$, form a linear basis in $(\n_+^1)^*$. The element $p_1$ of
$\ab_-$, which sends $d u^{(m)}_{i,\bar{1}}$ to $d u^{(m+1)}_{i,\bar{1}}$,
maps $(\n_+^m)^*$ isomorphically to $(\n_+^{m+1})^*$. Hence the vectors $d
u^{(m)}_{i,\bar{1}}$ are linearly independent. Therefore the functions
$u_i^{(n)}, i=1,\ldots,l, n\geq 0$, are algebraically independent and hence
$\C[U]$ embeds into $\C[Y]$. By definition, the function $u_i^{(n)}$ has
degree $-n$ with respect to the $\Z$--grading defined by the vector field
$\De^R$. Hence the character of $\C[U]$ is given by $$\on{ch} \C[U] =
\on{tr} q^{\Delta^R} = \prod_{n>0} (1-q^n)^{-l}.$$

On the other hand, the image of the adjoint action of $\ab_+$ coincides
with $\on{Im} p \simeq \n_+/\ab_+$. Recall that the exponential map $\exp:
\n_+ \arr N_+$ is an isomorphism. From Campbell-Baker-Hausdorff formula
we derive that any element of $N_+$ can be uniquely presented as the
product of an element of $A_+$ and an element $\exp x$ of $N_+$, where $x
\in \on{Im} p \subset \n_+$. Therefore $Y$ is isomorphic to $\n_+/\ab_+$.
Hence with respect to the $\Z$-grading defined by the vector field
$\Delta$, the space $\C[Y]$ of algebraic functions on $Y$ is a free
polynomial algebra with $l$ generators of each negative degree. Hence its
character coincides with the character of $\C[U]$. But since $\C[U]$ is
embedded into $\C[Y]$, $\C[U] \simeq \C[Y]$, and Proposition follows.
\end{pf}

We can now identify the algebra $\C[U]$ with the algebra of differential
polynomials $\pi_0$ from the previous section; the operator $p_1$ gets
identified with $\pa$.

\begin{thm}    \label{iso}
{\em $\C[Y]$ and $\pi_0$ are isomorphic as $\n_+$-modules.}
\end{thm}

\begin{pf} We have $$e_i = \sum_{1\leq j\leq l, n\geq 0} C_{i,j}^{(m)}
\fp{j}{m},$$ where $C_{i,j}^{(m)}$ are certain polynomials in $u_i^{(n)}$.
By definition of $u_i^{(0)}$ and formula \eqref{key}, we have
\begin{equation}    \label{ui}
[e_i,p_1] = -u_i^{(0)} e_i.
\end{equation}
Since $$p_1 = \sum_{1\leq i\leq l, n\geq 0} u_i^{(m+1)} \fp{i}{m},$$ from
formula \eqref{ui} we find the recurrence relations for the coefficient of
$\pa/\pa u_j^{(m-1)}$ in the vector field $e_i$: $$C_{i,j}^{(m)} =
-u_i^{(0)} C_{i,j}^{(m-1)} + p_1 \cdot C_{i,j}^{(m-1)}.$$ These recurrence
relations coincide with the recurrence relations for the coefficients of
the vector fields $Q_i$ on $U$. We also have, according to formulas
\eqref{property} and \eqref{nov}, $e_i
\cdot u_j^{(0)} = - (\al_i,\al_j),$ therefore $C_{i,j}^{(0)} = -
(\al_i,\al_j)$. Formulas \eqref{qi} and \eqref{rec} then show that the
vector field $e_i$ coincides with $Q_i$ and Proposition follows.
\end{pf}

\begin{rem} This result was proved by other methods in \cite{monte},
Propositions 3.1.10 and 3.2.5. Note that \propref{serre} follows from
it.\qed
\end{rem}

\begin{rem} Consider the homogeneous space $\widetilde{Y} = B_+/A_+$.
Since $B_+ = N_+ \times H$, we have: $\widetilde{Y} = Y \times H$. The
space of regular functions on $H$ can be identified with $\oplus_{\la \in
\La} \C e^{\bar{\la}}$ with multiplication $e^{\bar{\la}} \cdot
e^{\bar{\mu}} = e^{\bar{\la} + \bar{\mu}}$. Here $\La$ is the root lattice
in $\h^*$. Therefore, by \propref{homom}, the algebra $\C[\widetilde{Y}]$
of regular functions on $\widetilde{Y}$ is isomorphic to $\oplus_{\la \in
\La}\pi_\la$.

Define an action of the generator $h_i$ of the Cartan subalgebra $\h$ of
$\g$ on $\oplus_{\la \in \La} \pi_\la$ as $-\pa/\pa \phi_i$ and the
action of the generator $e_i$ of $\n_+ \in \g$ on $\oplus_{\la \in \La}
\pi_\la$ as $\q_i$ given by formula \eqref{borel}. One easily checks that
this defines an action of the Borel subalgebra $\goth{b}_+ = \h \oplus
\n_+$ of $\g$ on $\oplus_{\la \in \La} \pi_\la$. From \thmref{iso} we
derive the following result.
\end{rem}

\begin{cor}    \label{coriso}
$\C[\widetilde{Y}]$ and $\oplus_{\la \in \La} \pi_\la$ are isomorphic as
$\goth{b}_+$--modules.
\end{cor}

It is known that the group $B_+$ is equipped with a Lie-Poisson structure
\cite{drinfeld,drinfeld1}. On the other hand, via the isomorphism of
\corref{coriso}, $B_+/A_+$ is equipped with a generalized Hamiltonian
structure described in \secref{hf}. It would be interesting to connect
these two structures.

\section{BGG resolution and proof of Theorem 1}    \label{prone}
In this section we will show that the space of Toda integrals is isomorphic
to $\ab_+^*$. For that we will need the dual of the
Bernstein-Gelfand-Gelfand (BGG) resolution
\cite{bgg,rocha} and results of \secref{geocon}.

Recall that the dual of the BGG resolution is a complex $B^*(\g) =
\oplus_{j\geq 0} B^j(\g)$, where $B^j(\g) = \oplus_{l(w)=j} M^*_{w(\rho) -
\rho}$. Here $M^*_\la$ is the module contragradient to the Verma module of
highest weight $\la$, and the differentials of the resolution commute with
the action of $\g$. The $0$th cohomology of $B^*(\g)$ is one-dimensional
and all higher cohomologies of $B^*(\g)$ vanish, so that $B^*(\g)$ is an
injective resolution of the trivial representation of $\n_+$.

Using \propref{hom}, one can explicitly construct the differentials of the
dual BGG resolution.

It is known that for each pair of elements of the Weyl group, such that $w
\prec w'$, there is a singular vector $P_{w,w'} \cdot v_{w(\rho) - \rho}$
in $M_{w(\rho) - \rho}$ of weight $w'(\rho) - \rho$. By \propref{hom}, this
vector defines the homomorphism $\bar{P}_{w,w'}^L: M^*_{w(\rho) - \rho}
\arr M^*_{w'(\rho) - \rho}$.

It is possible to normalize all $P_{w,w'}$'s in such a way that
$P_{w'_1,w''} P_{w,w_1'} = P_{w_2',w''} P_{w,w_2'}$ for any quadruple of
elements of the Weyl group, satisfying $w \prec w_1', w_2'
\prec w''$. Then we obtain: $\bar{P}^L_{w,w_1'} \bar{P}^L_{w'_1,w''} =
\bar{P}^L_{w,w_2'} \bar{P}^L_{w_2',w''}$.

The differential $\delta^j: B^j(\g) \arr B^{j+1}(\g)$ of the BGG complex
can be written as follows: $$\delta^j = \sum_{l(w)=j,l(w')=j+1,w\prec w'}
\epsilon_{w,w'} \, \bar{P}_{w,w'}^L,$$ where $\epsilon_{w,w'} = \pm 1$ are
chosen as in \cite{bgg,rocha}.

By construction the right action of the Lie algebra $\g$ on this complex
commutes with the differentials. Therefore we can take the subcomplex of
invariants $F^*(\g) = B^*(\g)^{\ab_+^R}$ with respect to the action of the
Lie algebra $\ab_+^R$.

\begin{lem} The cohomology of the complex $F^*(\g)$ is isomorphic to the
exterior algebra $\bigwedge^*(\ab_+^*)$.
\end{lem}

\begin{pf} We have: $F^*(\g) = \oplus_{j\geq 0} F^j(\g)$, with
$$F^j(\g) = \oplus_{l(w)=j} \pi^{(w(\rho) - \rho)},$$ where $\pi^{(w(\rho)
- \rho)}$ denotes the space of $\ab_+$-invariants of the module
$M^*_{w(\rho) - \rho}$. As $\n_+$-modules all $\pi^{(w(\rho) - \rho)}$ are
isomorphic to $\pi_0$. Since $B^*(\g)$ is an injective resolution of the
trivial representation of $\n_+$, it is also an injective resolution of the
trivial representation of $\ab_+$. Therefore the cohomology of the complex
$F^*(\g)$ is nothing but $H^*(\ab_+,\C)$, which is $\bigwedge^*(\ab_+^*)$,
since $\ab_+$ is abelian.
\end{pf}

\begin{rem} The cohomology of the complex $F^*(\g)$ is also isomorphic to
$H^*(\n_+,\pi_0)$, cf. \cite{monte}, Proposition 2.4.5. The isomorphism
$H^*(\n_+,\pi_0) \simeq H^*(\ab_+,\C)$ follows from Shapiro's lemma.\qed
\end{rem}

The action of the Lie algebra $\ab_-^R$ on $B^*(\g)$ gives rise to an
$\ab_-$-action on the complex $F^*(\g)$. We know that if $x$ is an element
of $\ab_-$, then each of the functions $F_i(x)$ is invariant under the
$\ab_+^R$-action. Denote by $\bar{F}_i(x)$ the corresponding function on
$N_+/A_+$, $\bar{F}_i(x) \in \pi_0$. The action of $x$ on $\pi^{(\la)}$,
where $\la = \sum_{0\leq i\leq l} \la_i \al_i$, is given by the first order
differential operator
\begin{equation}    \label{operator}
x^R + \sum_{0\leq i\leq l} \la_i \bar{F}_i(x).
\end{equation}
By construction, this action commutes with the differentials of the complex
$F^*(\g)$.

Hence this action defines an action of $\ab_-$ on the cohomologies of the
complex $F^*(\g)$.

\begin{lem}    \label{action}
The action of $\ab_-$ on the cohomologies of the complex $F^*(\g)$ is
trivial.
\end{lem}

\begin{pf}
Let $\Omega^*(\n_+)$ be the de Rham complex of the big cell $X$ of the flag
manifold. This complex is an injective resolution of the trivial
representation of $\n_+$, which is isomorphic to the tensor product of the
space of functions on $X$ and the exterior algebra
$\bigwedge^*(\n_+^*)$. The Lie algebra $\g$ infinitesimally acts on
$\Omega^*(\n_+)$ from the right by vector fields, and this action commutes
with the differentials of the complex.

Let $\Omega^*(\ab_+)$ be the de Rham complex of the Lie group $A_+$. This
is an injective resolution of the trivial representation of $\ab_+$, which
is isomorphic to the tensor product of the space of functions on $A_+$ and
the exterior algebra $\bigwedge^*(\ab_+^*)$. The embedding $\ab_+
\arr \n_+$ induces a surjective homomorphism $\rho: \Omega^*(\n_+)
\arr \Omega^*(\ab_+)$. The corresponding homomorphism of $\ab_+$-invariants
$\bar{\rho}: \Omega^*(\n_+)^{\ab_+} \arr \Omega^*(\ab_+)^{\ab_+} =
\bigwedge^*(\ab_+^*)$ induces an isomorphism of the cohomologies.

The Lie algebra $\ab_- \subset \g$ acts on $\Omega^*(\n_+)$. Let $\ab_-$
act trivially on $\Omega^*(\ab_+)$. Since $\ab_-$ commutes with $\ab_+$,
the map $\rho$ commutes with the action of $\ab_-$. Hence the map
$\bar{\rho}$ also commutes with the action of $\ab_-$. Therefore the action
of $\ab_-$ on the cohomologies of $\Omega^*(\n_+)^{\ab_+}$ coincides with
its action on the cohomologies of $\Omega^*(\ab_+)^{\ab_+}$ and hence is
trivial.

The complex $B^*(\g)$ is a subcomplex of $\Omega^*(\n_+)$
\cite{bgg,rocha}, and the embedding $B^*(\g) \arr \Omega^*(\n_+)$ commutes
with the action of $\g$. Since both $B^*(\g)$ and $\Omega^*(\n_+)$ are
injective resolutions of the trivial representation of $\ab_+$, the map
$B^*(\g)^{\ab_+} \arr \Omega^*(\n_+)^{\ab_+}$ induces an isomorphism on
cohomologies. But it is also an $\ab_-$--homomorphism, therefore the action
of $\ab_-$ on the cohomologies of the complex $B^*(\g)^{\ab_+} = F^*(\g)$
is trivial.
\end{pf}

According to the Lemma, $p_1 \in \ab_-$ acts trivially on the
cohomologies. We already know that its action on $\pi_0$ coincides with the
action of $\pa$.  Consider now the action of $p_1$ on
$\pi^{(\la)}$. According to formula \eqref{operator}, it is given by $\pa +
\sum_{0\leq i\leq l} \la_i u_i^{(0)}$. But this coincides with the action
of $\pa$ on $\pi_\la$, given by \eqref{pala}. Therefore $\pi^{(\la)} \simeq
\pi_\la$ with respect to the action of $\n_+$ and with respect to the
action of $p_1 \equiv \pa$.
\vspace{5mm}

\noindent {\em Proof of \thmref{span}}. Since $\pa = p_1$ commutes with the
differentials of the complex $F^*(\g)$, we can consider the double complex
\begin{equation}    \label{double}
\C \larr F^*(\g) \stackrel{\pm p_1}{\larr} F^*(\g) \larr \C.
\end{equation}
Here $\C \arr \pi_0 \subset F^*(\g)$ and $F^*(\g) \arr \pi_0 \arr \C$ are
the embedding of constants and the projection on constants, respectively.
We place $\C$ in dimensions $-1$ and $2$ of our complex, and $F^*(\g)$ in
dimensions $0$ and $1$.

In the spectral sequence, in which $\pm p_1$ is the $0$th differential,
the first term is the complex $\F^*(\g)[-1]$, where $$\F^j(\g) \simeq
\oplus_{l(w)=j} \F_{w(\rho) - \rho}.$$ Indeed, if $\la \neq 0$, then in the
complex $$\pi_\la \stackrel{p_1}{\larr} \pi_\la$$ the $0$th cohomology is
$0$, and the first cohomology is, by definition, the space $\F_\la$. If
$\la = 0$, then in the complex $$\C \larr \pi_0 \stackrel{p_1}{\larr}
\pi_0 \larr \C$$ the $0$th cohomology is $0$ and the first cohomology is,
by definition, the space $\F_0$.

In particular, $\F^0(\g) = \F_0$, $\F^1(\g) = \oplus_{0\leq i\leq l}
\F_{-\al_i}$, and the differential $\bar{\delta}^0: F^0(\g) \arr F^1(\g)$
is given by $\bar{\delta}^0 = \sum_{0\leq i\leq l} \Q_i$. By definition,
the $0$th cohomology of the complex $\F^*(\g)$ and hence the $1$st
cohomology of the double complex \eqref{double} is isomorphic to the space
of Toda integrals.

We can compute this cohomology, using the other spectral sequence
associated to our double complex. Since $H^*(F^*(\g)) \simeq
\bigwedge^*(\ab_+^*)$, we obtain in the first term the following complex
$$\C \larr
\wedge^*(\ab_+^*) \stackrel{\pm p_1}{\larr} \wedge^*(\ab_+^*) \larr \C.$$
By \lemref{action}, the action of $p_1$ on $\bigwedge^*(\ab_+^*)$ is
trivial and hence the cohomology of the double complex \eqref{double} is
isomorphic to $\bigwedge ^*(\ab_+^*)/\C
\oplus \bigwedge ^*(\ab_+^*)/\C[-1]$. In particular, we see that the space of
Toda integrals is isomorphic to $\ab_+^*$.

With respect to the $\Z$--grading on $\F_\la$, introduced in Sect. 2, the
differentials of the complex are homogeneous of degree $0$. Moreover, the
corresponding $\Z$-grading on cohomology coincides with the one induced by
the principal grading on $\ab_+$. Therefore the space of Toda integrals is
linearly spanned by elements $H_m$ of degrees $-m \in -I$.\qed

\begin{rem} In \cite{monte}, Theorems 3.1.11 and 3.2.6, we proved
\thmref{span} in the case when all exponents of $\g$ are odd and the Coxeter
number is even (this excludes $D_{2n}^{(1)}, E_6^{(1)}$, and
$E_8^{(1)}$). In this case the degrees of all elements $p_m$ are odd. The
statement of \lemref{action} follows from simple degree counting in this
case, since the image of a cohomology class of odd degree under the action
of an operator of odd degree should be of even degree and hence should
vanish. In particular, it follows that $\pa = p_1$ acts trivially on
cohomologies, and we can apply the proof of \thmref{span} above.
\qed
\end{rem}

\section{Vector fields corresponding to Toda integrals}
Let us explain how to construct a Toda integral starting from a class in
the first cohomology of the complex $F^*(\g)$.

Consider such a class ${\cal H} \in \oplus_{0\leq i\leq l} \pi_{-\al_i}$.
Since $\pa = p_1$ acts trivially on cohomologies of the complex $\F^*(\g)$,
$\pa {\cal H}$ is a coboundary, i.e. there exists such $h \in \pi_0$ that
$\bar{\delta}^0 \cdot h = \pa {\cal H}$.

By construction, the element $h$ has the property that $\q_i \cdot h \in
\pi_{-\al_i}$ is a total derivative for $i = 0,\ldots,l$.  But it itself is
not a total derivative, because otherwise ${\cal H}$ would also be a
trivial cocycle. Therefore, $\int h \neq 0$. But then $\int h$ is a KdV
hamiltonian, because by construction $\bar{\delta}^0 \cdot \int h = \int
\bar{\delta}^0 \cdot h = 0$ and hence $\Q_i \cdot \int h = 0$ for any
$i=0,\ldots,l$.

For $m \in I$ denote by $H_m \in \F_0$ the Toda integral, corresponding to
an element of $\ab_+^*$ of degree $-m$. Denote by $\eta_m$ the derivation
$\xi_{H_m}$. In particular, simple calculation shows that we can choose as
${\cal H}_1$ vector $\sum_{0\leq i\leq l} e^{-\phi_i}$. Then $\pa {\cal
H}_1 = - \sum_{0\leq i\leq l} u_i^{(0)} e^{-\phi_i}$ and $h_1 = \frac{1}{2}
\sum_{1\leq i\leq l} u_i^{(0)} u^{(0)i}$. Hence $\eta_1 = \pa$, by
\eqref{pa}.

Now $\eta_m$ is a vector field on $Y \simeq U$. On the other hand the right
infinitesimal action of the generator $p_m$ of the Lie algebra $\ab_-
\subset \g$ on $U$ also defines a vector field on $U$, which we denote by
$\mu_m$.

\begin{thm}    \label{osnovnoi}
{\em The vector field $\eta_m$ coincides with the vector field $\mu_m$ up
to a non-zero constant multiple for any $m \in I$.}
\end{thm}

Note that we have already established this for $m=1$. Indeed, we have just
shown that $\eta_1 = \pa$, and we already know that the action of $\pa$
coincides with the action of $p_1$.

\begin{cor} The Toda integrals commute with each other: $$\{ H_n,H_m \} =
0$$ in $\F_0$ for any $n, m \in I$.
\end{cor}

\begin{pf} Since $p_m, m \in I,$ lie in a commutative Lie algebra, they
commute with each other. So do the corresponding vector fields:
$[\mu_n,\mu_m]=0$. By \thmref{osnovnoi}, the same holds for the vector
fields $\eta_m, m \in I$: $[\eta_n,\eta_m]=0$. By formula \eqref{com},
injectivity of the map $\xi$ on $\F_0$, and the definiton of the
vector fields $\eta_m$, the corresponding Toda integrals also commute
with each other.
\end{pf}

Let us now compute the commutator $[Q_j,\eta_m]$. From the definition of
the Toda integrals and \eqref{com} we obtain: $$[\q_j,\eta_m] = [-\xi_{\int
e^{-\phi_j}},\xi_{H_m}] = -\xi_{\{ \int e^{-\phi_j},H_m \}} = 0,$$ since
$\{ \int e^{-\phi_j},H_m \} = \int (\Q_j
\cdot H_m) = 0$. Therefore
\begin{equation}    \label{main}
[Q_j,\eta_m] = -(\delta_j H_m) Q_j.
\end{equation}
Indeed, in contrast to the operator $\q_j$, which acts from $\pi_0$ to
$\pi_{-\al_j}$, the operator $Q_j$ acts from $\pi_0$ to itself. Hence this
commutator should be equal to $\Delta_m^j \cdot Q_j$, where $\Delta_m^j$ is
the difference between the actions of the operator $\eta_m$ on
$\pi_{-\al_j}$ and $\pi_0$. This difference is equal to $$\sum_{1\leq i\leq
l} \delta_i H_m \frac{\pa e^{-\phi_j}}{\pa \phi_i} = -\delta_j H_m.$$

\section{Proof of Theorem 2}
Our proof will be based on formula \eqref{main}, which turns out to be a
defining property for vector fields $\eta_m$.

We will say that a vector field $\al$ on $N_+$ satisfies property (P), if
it satisfies formula
\begin{equation}    \label{newkey}
[e_i^L,\al] = -F_i(\al) e_i^L, \quad \quad i=0,\ldots,l,
\end{equation}
where $F_i(\al)$ are certain functions on $N_+$. Clearly, if $\al$ and
$\beta$ satisfy property (P), so does their commutator. Thus, vector
fields, which satisfy property (P), form a Lie subalgebra of $\V$, which
we denote by $\Li$. According to \propref{P}, $\G$ is a Lie subalgebra of
$\Li$ and we prove in \propref{desc} of the Appendix that in fact $\Li
\simeq \G$.

Consider now the vector field $\eta_m$ corresponding to the Toda integral
$H_m$, for some $m \in I$. By formula \eqref{main}, it satisfies property
(P) on the homogeneous space $N_+/A_+$.

We want to show that there exists an $\ab_+^R$--invariant vector field
$\widetilde{\eta}_m$ on $N_+$ satisfying property (P), such that its
projection to $N_+/A_+$ coincides with $\eta_m$.

Let us define a trivial one-cocycle $f_m$ on the Lie algebra $\n_+$ with
coefficients in vector fields on $N_+/A_+$ by putting $f_m(x) =
[x,\eta_m]$. Any one-cocycle $f$ satisfies the relation
\begin{equation}    \label{relation}
f([x,y]) = [x,f(y)] - [y,f(x)].
\end{equation}
Since $\n_+$ is generated by $e_i, i=0,\ldots,l$, $f$ is uniquely
determined by its values on $f_m(e_i)$.

Now suppose that we have assigned to each $e_i$ an element of our
module. These elements are values of a one-cocycle $f$ on the $e_i$'s, if
and only if the value of $f$ on any of the Serre relations, inductively
constructed using \eqref{relation}, vanishes. For example, if we have the
relation $[e_i,e_j] = 0$ in $\n_+$ and $f$ is a one-cocycle, then the
relation $[e_i,f(e_j)] - [e_j,f(e_i)] = 0$ should hold.

Our cocycle $f_m$ has a specific form due to the formula
\eqref{main}: the value of $f_m$ on each $e_i$ is proportional to
$e_i$, i.e. equal to $e_i$ multiplied by a function. By induction, one can
show that in this case the value of $f_m$ on the Serre relation $$(\on{ad}
e_i)^{-a_{ij}+1} \cdot e_j = 0$$ is a linear combination
\begin{equation}    \label{lincomb}
h_i e_i + h_j e_j + h_{ij} [e_i,e_j] + \ldots + h_{i\ldots i j} (\on{ad}
e_i)^{-a_{ij}} \cdot e_j,
\end{equation}
where $h_\bullet$ are certain functions on $N_+/A_+$.

For example, if we have $f_m(e_i) = g_i e_i, f_m(e_j) = g_j e_j$, and the
relation $[e_i,[e_i,e_j]]=0$, then $$f_m([e_i,[e_i,e_j]]) = (e_j e_i g_i -
2 e_i e_j g_i) e_i + (e_i^2 g_j) e_j + (e_i g_i + 2 e_i g_j) [e_i,e_j].$$

The linear term of the vector field $(\on{ad} e_i)^k \cdot e_j$ is non-zero
and has degree $k+1$ (cf. \cite{monte}, proofs of Propositions 3.1.10 and
3.2.5). Therefore the linear terms of these vector fields with $k>0$ are
linearly independent from each other and from the linear terms of the
vector fields $e_i$ and $e_j$, which have degree $1$. The linear terms of
the latters are given by $-\pa_i^{(0)}$ and $-\pa_j^{(0)}$ and hence are
also linearly independent, if $\g$ is not of rank two. But if $\g$ is of
rank two, then one can check directly that $e_0$ and $e_1$ are linearly
independent at a generic point of $Y$. Thus we see that the vector fields
$e_i, e_j, [e_i,e_j],\ldots,(\on{ad} e_i)^{-a_{ij}} \cdot e_j$, are
linearly independent at a generic point of $Y$.

Vanishing of the linear combination \eqref{lincomb} of these vector fields
multiplied by certain functions then implies that each of the functions
$h_\bullet$ vanishes identically on $N_+/A_+$.

We now want to ``lift'' the one-cocycle $f_m$ of $\n_+$ with coefficients
in vector fields on $N_+/A_+$ to a one-cocycle $\widetilde{f}_m$ of $\n_+$
with coefficients in vector fields on $N_+$ with respect to the left
action. For the one-cocycle $f_m$ we have: $$f_m(e_i) = -(\delta_i H_m)
e_i^L,$$ by \eqref{main}.

So we want to put
\begin{equation}    \label{put}
\widetilde{f}_m(e_i) = -\epsilon^* (\delta_i H_m) e_i^L,
\quad \quad i=0,\ldots,l,
\end{equation}
where $\epsilon$ is the projection $N_+ \arr N_+/A_+$. For
$\widetilde{f}_m$ to be a one-cocycle, the value of $\widetilde{f}_m$ on
any of the Serre relations, inductively constructed using \eqref{relation},
must vanish. But this value is given by formula \eqref{lincomb}, where we
should replace each of the functions $h_\bullet$ by
$\epsilon^*(h_\bullet)$. Therefore the value on a Serre relation is equal
to $0$. Hence there exists a one-cocycle $\widetilde{f}_m$ of $\n_+$ with
coefficients in vector fields on $N_+$, which satisfies \eqref{put}.

The cohomology $H^i(\n_+^L,\V)$ vanishes for $i>0$, because as a module
over $\n_+^L$, $\V$ is dual to a free module, with $\n_+^R$ as the space of
invariants. In particular, $H^1(\n_+^L,\V) = 0$, and therefore there exists
a vector field $\widetilde{\eta}_m$ on $N_+$, such that $\widetilde{f}_m(x)
= [x,\widetilde{\eta}_m]$. In particular, we obtain:
\begin{equation}    \label{exist}
[e_i^L,\widetilde{\eta}_m] = -\epsilon^* (\delta_i H_m) e_i^L, \quad \quad
i=0,\ldots,l.
\end{equation}

Thus, the vector field $\widetilde{\eta}_m$ satisfies property (P). By
\propref{desc}, it lies in $\G$. Let $p^R$ be the vector field of the right
infinitesimal action of an element $p \in \ab_+ \subset \n_+$ on $N_+$. We
have $$[e_i^L,[p^R,\widetilde{\eta}_m]] = [[e_i^L,p^R],\widetilde{\eta}_m]
+ [p^R,[e_i^L,\widetilde{\eta}_m]] = 0,$$ for any $i=0,\ldots,l$, because
$[e_i^L,p^R] = 0$ and $$[p^R,[e_i^L,\widetilde{\eta}_m]] = - \left( p^R
\cdot \epsilon^* (\delta_i H_m) \right) e_i^L - \epsilon^* (\delta_i H_m)
[e_i^L,p^R] = 0,$$ since by definition $$x \cdot \epsilon^* (\delta_i H_m)
= 0 \quad \quad \forall x \in \ab_+^R.$$

Therefore, by \propref{P}, (a), $[p^R,\widetilde{\eta}_m] \in \n_+^R$. But
the degree of $\widetilde{\eta}_m$ is equal to $-m<0$, so that the degree
of the commutator with $p$ is equal to $-m+1 \leq 0$, whereas the degree of
any element of $\n_+^R$ should be positive. We conclude that
$[p^R,\widetilde{\eta}_m] = 0$.

But $\widetilde{\eta}_m \in \G$, and the only elements of $\G$, which
commute with $p^R$, are elements of $\ab^R$. Comparing degrees we see that
$\widetilde{\eta}_m$ coincides with the vector field $p_m^R$, corresponding
to a generator $p_m$ of $\ab_-$ of degree $-m$. This generator acts on
$N_+/A_+$ by the vector field $\mu_m$. By construction, $[\mu_m-\eta_m,e_i]
= 0$ for any $i=0,\ldots,l$. Hence $\mu_m-\eta_m$ is an $\n_+$--invariant
vector field on $N_+/A_+$.

The space of vector fields on $N_+/A_+$ as a module over $\n_+$ is
coinduced from the $\ab_+$--module $\n_+/\ab_+$. Therefore the space of
$\n_+$--invariant vector fields on $N_+/A_+$ is isomorphic to the space of
$\ab_+$--invariants of $\n_+/\ab_+$. The latter space is $0$, and hence
$\mu_m-\eta_m=0$. Therefore the vector field $\mu_m$ of the infinitesimal
right action of the element $p_m \in \ab_-$ on $N_+/A_+$ coincides with the
vector field $\eta_m$ of the $m$th Toda integral $H_m$.\qed

\section{Appendix}
The Lie algebra $\Li$ can be defined for an arbitrary Kac-Moody algebra
$\g$ as the Lie algebra of vector fields on the big cell of the flag
manifold, satisfying the relations \eqref{newkey}.

The Lie algebra $\Li$ preserves a certain geometric structure on the flag
manifold $F$. Denote by $P_i, i \in J$, the parabolic subgroup of $G$,
obtained by adjoining to the Borel subgroup $B_+$ the one-parameter
subgroup of the negative simple root generator $f_i$ of $\g$. Here $J$ is
the set of simple roots of $\g$. We have natural bundles: $F \arr G/P_i, i
\in J$. The fiber of such a bundle is a projective line. The tangent spaces
to the fibers of these bundles defines $|J|$ tangent directions at each
point of $F$. The Lie algebra $\Li$ consists of vector fields, which are
infinitesimal symmetries of this structure. It seems plausible that all
elements of ${\cal L}$ for $\g \neq \sw_2$ are restrictions to $N_+$ of
globally defined vector fields on the flag manifold $B_-\backslash G$.

Clearly, $\g$ itself is a Lie subalgebra of $\Li$. The knowledge of $\Li$
is important for understanding representation theory of $\g$. Let $\beta$
be an element of $\Li$ and $\la = \sum_{i \in J} \la_i \al_i$ be a weight
of the Cartan subalgebra of $\g$, where $\al_i, i \in J$, are the simple
roots of $\g$ and $\la_i$'s are complex numbers. Define a map from $\Li$ to
the Lie algebra of differential operators of the first order on the big
cell of $F$, which sends $\beta \in \Li$ to the sum of the vector field,
corresponding to $\beta$, and the function $\sum_{i \in J}^n \la_i
F_i(\beta)$, where the $F_i(\beta)$'s are defined by the relations
\eqref{newkey}. This is clearly a homomorphism of Lie algebras. As we already
mentioned above, the space of functions on the big cell of $F$ with respect
to this action of the Lie algebra $\g \subset \Li$ coincides with the
contragradient module to the Verma module with highest weight $\la$,
$M^*_\la$. Thus we obtain a structure of $\Li$-module on $M^*_\la$. Since
any irreducible module from the category ${\cal O}$ of $\g$ can be realized
as a submodule of some $M^*_\la$, we obtain a structure of $\Li$-module on
an arbitrary $\g$-module.

We do not know a description of $\Li$ for Kac-Moody algebras other than
finite-dimensional or affine.

\begin{prop} If $\g$ is a finite-dimensional simple Lie algebra other than
$\goth{sl}_2$, then the Lie algebra $\Li$ coincides with $\g$. For
$\g=\goth{sl}_2$ the Lie algebra $\Li$ coincides with the Lie algebra of
vector fields on the line.
\end{prop}

\begin{pf}
In the same way as in the proof of \propref{desc}, we reduce the problem to
the calculation of the cohomology $H^1(\n_+,\g)$. Since $\g$ is
finite-dimensional, we can use the Borel-Weil-Bott-Kostant theorem
\cite{b,k}, which gives: $$H^1(\n_+,\g) = \oplus_{i=1,\ldots,l}
\C_{s_i(\Lambda_{adj} +\rho) - \rho}.$$ Here $l$ is the rank of $\g$,
$\Lambda_{adj}$ is the highest weight of the adjoint representation, $\rho$
is the half-sum of the positive roots of $\g$, and $s_i, i=1,\ldots,l,$ are
the reflections from the Weyl group of $\g$. We denote by $\C_\la$ a
one-dimensional representation of the Cartan subalgebra of $\g$, on which it
acts according to its character $\la$.

If $\Li \neq \g$, then there should exist a vector field $\beta \in \Li$,
whose weight and hence the weights of the
corresponding functions $F_j(\beta)$ should be equal to $s_i(\Lambda_{adj}
+ \rho) - \rho$ for some $i=1,\ldots,l$ (cf. the proof of
\propref{desc}). If $\g$ is not $\goth{sl}_2$, all of the weights
$s_i(\Lambda_{adj} + \rho) - \rho$ are non-zero and non-negative. Therefore
$\beta$ can not satisfy property (P), because functions on the big cell can
only have negative or zero weights, and hence $\Li = \g$.

If $\g=\goth{sl}_2$, one can check by hand that all vector fields on the big
cell satisfy property (P).
\end{pf}

\begin{prop}    \label{desc}
If $\g$ is an affine algebra, $\Li$ is isomorphic to $\G$.
\end{prop}

\begin{pf} The Lie algebra $\g$ acts on $\V$ by commutation.
We have the exact sequence $$0 \larr \G \larr \Li \larr M \larr 0$$ of
$\g$-modules. We will show that the module $M$ belongs to the category
${\cal O}$ of modules over $\g$
\cite{bgg,rocha}. A module $L$ belongs to the category ${\cal O}$, if it
satisfies two properties: (1) the Cartan subalgebra $\h$ of $\g$ acts on
$L$ diagonally; (2) for any $x \in M$, $U(\n_+) \cdot x$ is
finite-dimensional. Let us show that these properties are satisfied for the
$\g$--module $M$.

But we have already seen that the first property is satisfied on the whole
Lie algebra $\V$. Recall that the Cartan subalgebra $\h$ maps to
$\V$. One easily checks that the adjoint action of the image of $\h$
defines a grading of $\V$ by the weights of $\h$, and that $\Li$, and hence
$M$, are graded Lie subalgebras of $\V$. Note that in $M$ only negative or
$0$ weights can occur. Indeed, by definition, the commutation relations in
$\V$ preserve the grading. Formula \eqref{newkey} then implies that the weight
of a vector field $x \in \Li$ is equal to the weight of each of the
functions $F_i(x)$. But functions on $X$ can only have negative or $0$
weights, and if all $F_i(x)$'s vanish, then $x \in \n_+^L \subset \G$ by
\propref{P}, (a).

To show that the second property is satisfied, consider an element $x$ of
$M$ of weight $\gamma$. If $y$ is an element of $\n_+$, then, since $y^R$
commutes with $\n_+^L$, we obtain: $$[[y^R,x],e_i^L] = - \left( y^R
\cdot F_i(x) \right) e_i^L.$$ As an $\n_+^R$-module, the space of
functions on $X$ is isomorphic to the dual of the free module with one
generator. The weight of each of $F_i(x)$ is equal to $\gamma$, therefore
any element of $U(\n_+^L)$ of weight greater than $-\gamma$ maps $x$ to a
vector field, which commutes with all the $e_i^L$'s and hence lies in
$\n_+^R \subset \G$. The subspace of $U(\n_+)$ of elements of weight less
than or equal to $-\gamma$ is finite-dimensional. Therefore $U(\n_+) \cdot
x$ is finite-dimensional mod $\G$.

Suppose that $M \neq 0$. Then it should contain a highest weight vector,
i.e. a vector field $\nu$, which satisfies $[\n_+^R,\nu] \in \G$. But then
$\C \nu
\oplus \G$ should be an extension of a trivial one-dimensional
$\n_+^R$-module by the $\n_+^R$-module $\G$. This extension must be
non-trivial, because otherwise we would be able to find $\nu' \in \C \nu
\oplus \G$, such that $[\n_+^R,\nu'] = 0$. But then by \propref{P}, (a),
$\nu' \in \n_+^L$, and then $\nu'$ can not satisfy property (P).

Thus, this extension should define a non-zero element in the group
$H^1(\n_+,\G)$. Recall that $\G = \g \times \Ve_-$, where $\g' = [\g,\g]/\C
K$. The cohomology $H^*(\n_+,\g')$ was computed in \cite{ffial}:
$H^i(\n_+,\g') \simeq H^{i-1}(\n_+,\C) \otimes H^1(\n_+,\g')$. The space
$H^1(\n_+,\g')$ is naturally identified with the Lie algebra $\Ve =
\C[t^k,t^{-k}] t\pa_t$ of vector
fields on the circle. A vector field $\delta$ defines a $1$-cocycle
$f_\delta$ on $\n_+$ with coefficients in $\g'$ by the formula $f_\delta(x)
= [\delta,x]$.

{}From the long exact sequence associated with the short exact sequence of
$\n_+$-modules $$0 \larr \g' \larr \G \larr \Ve_- \larr 0$$ we obtain:
$H^i(\n_+,\G) \simeq H^i(\n_+,\C) \otimes \Ve_+$, where $\Ve_+ = t^k
\C[t^k] t\pa_t$. In particular, $H^1(\n_+,\G) \simeq \Ve_+$. But then the
vector field $\nu$, defining the extension, should have a positive
weight. Therefore the functions $F_i(\nu), i=0,\ldots,l,$ in formula
\eqref{newkey} should also have positive weights and hence they must
vanish. But then $\nu \in \n_+^R \subset \G$, by \propref{P}, (a), and so
$M=0$.
\end{pf}

It is interesting to notice that according to \propref{desc}, the flag
manifold of an affine algebra ``knows'' that a half of the Lie algebra of
vector fields on the circle acts on $\g$ by exterior automorphisms and that
this action lifts to any highest weight $\g$--module. Thus, although the
flag manifold comes from the Kac-Moody definition of $\g$, it also contains
some information about $\g$ as the central extension of a loop
group. Finding $\Li$ for general Kac-Moody algebras may therefore shed
light on their possible ``hidden symmetries''.

\vspace{10mm}
\noindent {\bf Acknowledgments.} We would like to thank T. Inami, M.
Kashiwara, and T. Miwa for their warm hospitality during our visit to Kyoto
University in the Summer of 1993, when the main part of this work had been
done.

We also thank M. Kashiwara for a useful discussion and S. Kumar for
correspondence.

We gratefully acknowledge financial support from Kyoto University.  The
research of the second author was also supported by a Junior Fellowship
from the Society of Fellows of Harvard University and by NSF grant
DMS-9205303.

\end{document}